\documentclass[twocolumn,english,showpacs,preprintnumbers,amsmath,amssymb,floatfix]{revtex4}
\usepackage[T1]{fontenc}
\usepackage[latin9]{inputenc}
\usepackage{color}
\usepackage{array}
\usepackage{amstext}
\usepackage{graphicx}
\usepackage{esint}

\makeatletter

\@ifundefined{textcolor}{}
{%
 \definecolor{BLACK}{gray}{0}
 \definecolor{WHITE}{gray}{1}
 \definecolor{RED}{rgb}{1,0,0}
 \definecolor{GREEN}{rgb}{0,1,0}
 \definecolor{BLUE}{rgb}{0,0,1}
 \definecolor{CYAN}{cmyk}{1,0,0,0}
 \definecolor{MAGENTA}{cmyk}{0,1,0,0}
 \definecolor{YELLOW}{cmyk}{0,0,1,0}
 }

\@ifundefined{definecolor}
 {\usepackage{color}}{}
\@ifundefined{definecolor}
 {\usepackage{color}}{}
\makeatother

\makeatother

\usepackage{babel}

\begin{document}

\title{Energy spectrum of bottom- and charmed-flavored mesons from polarized  
 top quark decay $t(\uparrow)\rightarrow W^+ + B/D+X$ at ${\cal O}(\alpha_s)$}

\author{S. Mohammad Moosavi Nejad $^{a,b}$}

\email{mmoosavi@yazduni.ac.ir}

\affiliation{$^{(a)}$Faculty of Physics, Yazd University, P.O. Box
89195-741, Yazd, Iran}

\affiliation{$^{(b)}$School of Particles and Accelerators,
Institute for Research in Fundamental Sciences (IPM), P.O.Box
19395-5531, Tehran, Iran}

\date{\today}

\begin{abstract}

We consider the decay of a polarized top quark into a stable $W^+$ boson and  charmed-flavored (D) 
or bottom-flavored (B) hadrons, via $t(\uparrow)\rightarrow W^++D/B+X$. We study the angular distribution of the scaled-energy 
of B/D-hadrons at next-to-leading order (NLO) considering the contribution of bottom  and  gluon fragmentations
into the heavy mesons B and D. To obtain the energy spectrum of B/D-hadrons we present our analytical expressions for the 
parton-level differential decay widths of $t(\uparrow)\rightarrow b+W^+(+g)$ at NLO. Comparison of our predictions 
with data at the LHC enable us to test the universality and scaling violations of the B- and D-hadron 
fragmentation functions (FFs). These can also be used to determine the polarization
states of top quarks and since the energy distributions depend on the ratio $m_W /m_t$ we advocate the use of such angular decay measurements for the
determination of the top quark's mass.
\end{abstract}

\maketitle

\section{INTRODUCTION}

The top quark is the heaviest particle of the Standard Model (SM) of elementary particle physics
and its short lifetime  implies that it
decays before hadronization takes place, therefore it retains
 its full polarization content when it decays. This allows one to study the top spin
state  using the angular distributions of its decay products. 
Highly polarized top quarks  become available at hadron
colliders through single top production processes, which occurs at the $33\%$ level
of the $t\bar t$  pair production rate \cite{Mahlon:1996pn}. Near maximal and
minimal values of top quark polarization at a linear $e^+e^-$ collider can be achieved in 
$t\bar{t}$ production by tuning the longitudinal polarization of the beam polarization \cite{Groote:2010zf} so that
a polarized linear $e^+e^-$ collider may be viewed as a copious source of close to zero
and close to $100\%$ polarized top quarks.  
A first study of polarization in $t\bar{t}$ events was performed
by the $D0$ collaboration \cite{Abazov:2012oxa} which showed good agreement between
the SM prediction and data.\\
The top quark total width, $\Gamma_t$, is proportional to the third power of its mass and is much larger than the 
typical QCD scale $\Lambda_{QCD}$. Therefore it enables us to treat the top quark almost as a free particle 
and to apply perturbative methods to evaluate the quantum corrections to its decay process.\\
The Large Hadron Collider (LHC) is a formidable top factory which is designed to produce about 90
million top quark pairs per year of running at design c.m. energy $\sqrt{S}=14$ TeV and design luminosity 
$10^{34} cm^{-2}s^{-1}$ in each of the four experiments \cite{Moch:2008qy}. This will allow one to specify the properties of the
top quark, such as its total decay width $\Gamma_t$, mass $m_t$ and branching fractions to high accuracy.
The theoretical aspects of top quark physics at the LHC are summarized in \cite{Bernreuther:2008ju}.\\
Due to the element $|V_{tb}|=0.999152$ of the CKM \cite{Cabibbo:1963yz} quark mixing matrix, top quarks almost exclusively decay to bottom
quarks, via $t\rightarrow bW^+$ followed by bottom quarks hadronization, $b\rightarrow H_b+X$, before b-quarks decay.
Here, $H_b$ stands for the observed final state hadron fragmented from the b-quark so that its production process 
regarded as the nonperturbative aspect of $H_b$-hadron formation from top decays. Of particular interest are
 the distribution in the scaled $H_b$-hadron energy $x_H$ in the top quark 
rest frame as reliably as possible. 
In Ref.~\cite{Kniehl:2012mn} we have calculated the doubly differential distribution
$d^2\Gamma/(dx_Bd\cos\theta)$ of the partial width of the decay $t\rightarrow bW^+\rightarrow Be^+\nu_e+X$
where $\theta$ is the decay angle of the positron in the W-boson rest frame and
$x_B$ is the scaled-energy of bottom-flavored hadrons B.
In Ref.~\cite{MoosaviNejad:2011yp}, we also evaluated the first order QCD corrections to 
the energy distribution of B-hadrons 
from  the decay of an unpolarized top quark into a charged-Higgs boson, via $t\rightarrow bH^+\rightarrow BH^++X$, in the theories 
beyond-the-SM with an extended Higgs sector. In Ref.~\cite{Ali:2011qf}, is mentioned that 
a clear separation between the decay modes $t \rightarrow bW^+$ and $t \rightarrow bH^+$ can be achieved 
in both the $t\bar{t}X$ pair production and the  $t/\bar{t}X$ single top production at the LHC. \\
Since  bottom quarks hadronize before they decay, the 
particular purpose of this paper is the evaluation of the NLO QCD corrections to the energy distribution of 
charmed-flavored (D) and 
bottom-flavored (B) hadrons from the decay of a polarized top quark into a bottom quark, via
$t(\uparrow)\rightarrow W^++b(\rightarrow B/D+X)$ where D stands for one of the mesons $D^0$, $D^+$ and $D^{\star +}$. 
These measurements will be important to deepen our understanding of the 
nonperturbative aspects of D- and B-mesons  formation which are described 
by realistic and nonperturbative fragmentation functions (FFs). The $b\rightarrow B$ FF is obtained through a global fit to
$e^+e^-$ data from CERN LEP1 and SLAC SLC in Ref.~\cite{Kniehl:2008zza} and the $b\rightarrow
 D^0/D^+/D^{\star +}$ FFs are determined in \cite{Kneesch:2007ey} 
by fitting the $e^+e^-$ experimental data from the BELLE, CLEO, OPAL and ALEPH collaborations.
As was demonstrated in \cite{Kniehl:2012mn}, the finite-$m_b$ corrections are rather small and thus 
to study the distributions in the heavy meson scaled-energy ($x_B$ for B-meson and $x_D$ for D-mesons), we employ  
the massless scheme or  zero-mass variable-flavor-number (ZM-VFN) scheme \cite{jm} in the top quark rest frame, where 
the zero mass parton approximation is also applied  to the bottom quark. The non-zero value of the b-quark mass only enter
through the initial condition of the nonperturbative FF.

This paper is structured as follows.
In Sec.~\ref{sec:one}, we introduce the angular structure of differential decay width by defining 
the technical details of our calculations.
In Sec.~\ref{sec:two}, our analytic results for the ${\cal O}(\alpha_s)$ QCD corrections to the 
angular distributions of partial decay rates are presented.
In Sec.~\ref{sec:three}, we present our numerical analysis.
In Sec.~\ref{sec:four},  our conclusions are summarized.

\boldmath
\section{ANGULAR DECAY DISTRIBUTION}
\label{sec:one}
\unboldmath

\begin{figure}
\begin{center}
\includegraphics[width=0.7\linewidth]{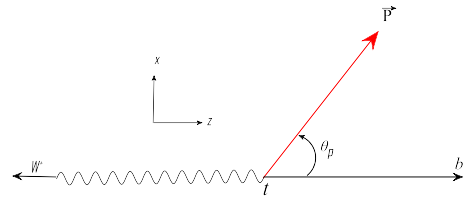}
\caption{\label{lo}%
Definition of  the  polar angle $\theta_P$ in the top quark rest frame.
$\vec{P}$ is the polarization vector of the top quark.}
\end{center}
\end{figure}

In this section we explain the ${\cal O}(\alpha_s)$ radiative corrections to the partial decay rate
$t(\uparrow)\rightarrow W^++b$.
The dynamics of the current-induced $t\rightarrow b$ transition is
depicted in the hadron tensor $H^{\mu\nu}$ which is introduced by
\begin{eqnarray}\label{tensor}
H^{\mu\nu}&=&(2\pi)^3\sum_{X_b}\int d\Pi_f\delta^4(p_t-p_W-p_{X_b})\nonumber\\
&&\times \frac{1}{2m_t}<t(p_t,s_t)|J^{\nu+}|X_b><X_b|J^{\mu}|t(p_t,s_t)>,\nonumber\\
\end{eqnarray}
where $d\Pi_f$ refers to the Lorentz-invariant phase space factor. In the SM the weak current is 
given by $J^{\mu}\propto \bar{q_b}\gamma^{\mu}(1-\gamma_5)\bar{q_t}$. 
 Since we are not summing over the top quark spin the hadron tensor $H^{\mu\nu}$
also depends on the top spin $s_t$. In this work we shall be
concerned  only with two types of intermediate state in Eq.~(\ref{tensor}), i.e. $|X_b>=|b>$ for Born
term and ${\cal O}(\alpha_s)$ one-loop contributions and $|X_b>=|b+g>$ for ${\cal O}(\alpha_s)$
tree graph contribution.
 The decay $t(\uparrow)\rightarrow W^++b$ is analyzed in the rest frame of the top quark (Fig.~\ref{lo}) where the
three-momentum of the $b$-quark points into the direction of the positive z-axis. The polar angle
$\theta_P$ is defined as the angle between the top quark polarization vector $\vec{P}$  and the z-axis.
We shall closely follow the notation of \cite{Kniehl:2012mn}  where we discussed the ${\cal O}(\alpha_s)$ radiative corrections 
to the partial decay rate of unpolarized top quarks.

The angular distribution of the top quark differential decay width $d\Gamma/dx$ is given by the following simple expression
to clarify the correlations between the top quark decay products and the top spin
\begin{eqnarray}\label{widthdefine}
\frac{d^2\Gamma}{dx_b d\cos\theta_P}=\frac{1}{2}(\frac{d\Gamma^{unpol}}{dx_b}+P\frac{d\Gamma^{pol}}{dx_b}\cos\theta_P),
\end{eqnarray}
where  we have defined the  b-quark scaled-energy as
\begin{eqnarray}\label{scale}
x_b=\frac{2E_b}{m_t(1-\omega)},
\end{eqnarray}
that $\omega$ being $\omega=m_W^2/m_t^2$. 
 Neglecting the b-quark mass one has $0\leq x_b \leq 1$. 
 In Eq.~(\ref{widthdefine}), P is the magnitude of the top quark polarization with $0\leq P\leq 1$ such that
$P=0$ corresponds to an unpolarized top quark and $P=1$ corresponds to $100\%$ top quark polarization.
$d\Gamma^{unpol}/dx_b$ and $d\Gamma^{pol}/dx_b$ stand for the unpolarized and polarized differential rates, respectively. 
In the following we discuss the technical details of our calculation for the ${\cal O}(\alpha_s)$ radiative 
corrections to the tree-level decay rate of $t(\uparrow)\rightarrow b+W^+$ using dimensional regularization.

\boldmath
\subsection{BORN TERM RESULTS}
\label{sec:two-1}
\unboldmath

The Born term tensor is calculated by squaring the Born term amplitude which is given by
\begin{eqnarray}
M_0=\frac{-e V_{tb}}{2\sqrt{2}\sin\theta_W}\bar{u}(p_b, s_b)\gamma^{\mu}(1-\gamma_5)
u(p_t, s_t)\epsilon^{\star}_{\mu}(p_W, \lambda),\nonumber\\
\end{eqnarray}
where $\epsilon^\mu$ stands for the W-boson polarization vector, the angle $\theta_W$
is the weak mixing angle and we take $\sin^2\theta_W=0.23124$ \cite{Caso:1998tx}, $s$ and $p$ stand for the spin and four-momenta 
of particles, respectively. The squared Born amplitude is expressed as
\begin{eqnarray}\label{bornamplitude}
|M_0|^2&=&\sum_{s_b, s_W}M_0.M_0^{\dagger}=\frac{e^2}{2\sin^2\theta_W}[p_b.p_t-m_t p_b.s_t+\nonumber\\
&&\frac{2}{m^2_W}(p_b.p_W)(p_t.p_W)-2\frac{m_t}{m_W^2}(p_b.p_W)(p_W.s_t)],\nonumber\\
\end{eqnarray}
where we replaced $\sum_{s_t}u(p_t, s_t)\bar{u}(p_t, s_t)=(\displaystyle{\not}p_t+m_t)$ in the unpolarized Dirac string by
 $u(p_t, s_t)\bar{u}(p_t, s_t)=1/2(1-\gamma_5\displaystyle{\not}s_t)(\displaystyle{\not}p_t+m_t)$ 
in the polarized state.\\
Considering Fig.~\ref{lo}, we parameterize the four-momenta and the polarization vector in the top rest frame as
\begin{eqnarray}
p_b&=&(E_b, 0, 0, E_b), \quad\quad\quad p_W=(E_W, 0, 0, -E_b),\nonumber\\
&&\hspace{-0.5cm} s_t=P(0, \sin\theta_P\cos\phi_P, \sin\theta_P \sin\phi_P, \cos\theta_P),\nonumber\\
\end{eqnarray}
where the parameter P is the degree of polarization. Therefore, the LO amplitude squared reads
\begin{eqnarray}\label{amplitude}
|M_0|^2=\frac{\pi \alpha}{\sin^2\theta_W}\frac{1-\omega}{\omega}m_t^2[1+2\omega+P(1-2\omega)\cos\theta_p].\nonumber\\
\end{eqnarray}
Here  $\pi\alpha/\sin^2\theta_W$ is related to the Fermi's constant $G_F$ as $\sqrt{2}m_W^2G_F$.\\
The decay rate of a particle with a mass $m$ and momentum $p$ into a given final state of particles $(p_1, p_2, \cdots, p_n)$
is
\begin{eqnarray}\label{gamma}
d\Gamma&=&\nonumber\\
&&\hspace{-1cm}\frac{1}{2m}\prod_{i=1}^n\frac{d^3p_i}{(2\pi)^3}\frac{1}{2E_i}(2\pi)^4\delta^4(p-\sum_{i=1}^n p_i)|M(m\rightarrow\{p_i\})|^2,\nonumber\\
&&\hspace{-1cm}=\frac{1}{2m}\frac{1}{(2\pi)^{3n-4}}dR_n(p, p_1,p_2,\cdots,p_n)|M(m\rightarrow\{p_i\})|^2,\nonumber\\
\end{eqnarray}
that in the two-body phase space (e.g. $t\rightarrow b+W^+$), using the following relation
\begin{eqnarray}
\int\frac{d^3\vec{p}}{2E}=\int d^4 p\delta(p^2-m^2)\Theta(E),
\end{eqnarray}
one has
\begin{eqnarray}\label{phase}
dR_2(p_t, p_b, p_W)=\frac{1-\omega}{8}(2\pi d\cos\theta_P).
\end{eqnarray}
Substituting (\ref{amplitude}) and (\ref{phase}) into (\ref{gamma}), one has
\begin{eqnarray}\label{bornfinal}
\frac{d\Gamma_0}{d\cos\theta_P}=\frac{1}{2}\{\Gamma^{unpol}_0+P\Gamma^{pol}_0\cos\theta_P\},
\end{eqnarray}
where  the polarized and unpolarized Born term  rates read
\begin{eqnarray}
\Gamma^{unpol}_0=\frac{m_t^3 G_F}{8\sqrt{2}\pi}(1+2\omega)(1-\omega)^2,
\end{eqnarray}
and 
\begin{eqnarray}
\Gamma^{pol}_0=\frac{m_t^3 G_F}{8\sqrt{2}\pi}(1-2\omega)(1-\omega)^2.
\end{eqnarray}
The polarization asymmetry $\alpha_W$ is defined by $\alpha_W=\Gamma_0^{pol}/\Gamma_0^{unpol}$ 
which is simplified to $\alpha_W=(1-2\omega)/(1+2\omega)=0.396$ if we set $m_W=80.399$ GeV and $m_t=172.9$ GeV \cite{Nakamura:2010zzi}.

\boldmath
\subsection{ONE-LOOP CONTRIBUTION}
\label{sec:two-2}
\unboldmath

The virtual one-loop  contributions  to the fermionic (V-A) transitions have a long history. Since at
the one-loop level,   QED and QCD have the same structure then the history   
even dates back to QED times. 
In this section we will investigate the  one-loop corrections and describe the method 
applied to extract the singularities at zero-mass scheme.
In the ZM-VFN scheme, where $m_b=0$ is set right from the start, both the soft and collinear singularities are regularized 
by dimensional regularization in $D=4-2 \epsilon$ space-time dimensions to become single poles in $\epsilon (0<\epsilon\leq 1)$. These 
singularities are subtracted at factorization scale $\mu_F$ and absorbed into the bare FFs according to the
modified minimal-subtraction $(\overline{MS})$ scheme. This renormalizes the FFs and creates in $d\Gamma/dx_b$ finite terms 
including the term $\alpha_S\ln(\mu_F^2/m_t^2)$ which are rendered perturbatively small by choosing $\mu_F={\cal O}(m_t)$.\\
\begin{figure}
\begin{center}
\includegraphics[width=1\linewidth]{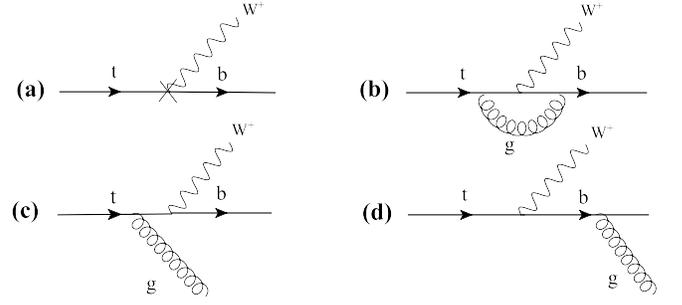}
\caption{\label{feynmandiagrams}%
Virtual (a, b) and real gluon (c, d) contributions  to $t(\uparrow)\rightarrow b+W^+$ at NLO.}
\end{center}
\end{figure}
 The virtual  contributions exhibit both the infrared (IR) and ultraviolet (UV) singularities which are regularized in D-dimensions.
 To evaluate the one-loop contributions to $t(\uparrow)\rightarrow b+W^+$ we consider the Feynman diagrams 
drawn in Fig.~\ref{feynmandiagrams}.
The renormalized amplitude of the virtual corrections can be written as
\begin{eqnarray}
M_{loop}=\frac{-e}{2\sqrt{2}\sin\theta_W}\epsilon^{\star}_{\mu}(p_W)\bar{u}(p_b, s_b)
\{\Lambda_\mu+\delta\Lambda_\mu\}u(p_t, s_t).\nonumber\\
\end{eqnarray}
Considering Fig.~\ref{feynmandiagrams}a, the counter term of the vertex is given by $\delta\Lambda_\mu=(\delta Z_b/2+\delta Z_t/2)\gamma_{\mu}(1-\gamma_5)$
where the wave-function renormalization constants of the top ($\delta Z_t$) and bottom ($\delta Z_b$) can be found in \cite{MoosaviNejad:2011yp}.
From Fig.~\ref{feynmandiagrams}b, for the one-loop vertex correction one has
\begin{eqnarray}\label{oneloop}
\Lambda_\mu=\frac{C_F\alpha_S}{4i\pi^3}\int d^4 p_g\frac{\gamma^\beta(\displaystyle{\not}p_b+\displaystyle{\not}p_g)
\gamma_\mu(1-\gamma_5)(\displaystyle{\not}p_t+m_t)\gamma_\beta}{p_g^2(p_b+p_g)^2(p_t^2-m_t^2)},\nonumber\\
\end{eqnarray}
where $C_F=4/3$ stands for the color factor and $p_g$ refers to the gluon four-momenta. The integral (\ref{oneloop}) is both ir- and uv-divergent that we 
use dimensional regularization to extract singularities taking the replacement 
\begin{eqnarray}
\int\frac{d^4p_g}{(2\pi)^4} \rightarrow \mu^{4-D}\int\frac{d^Dp_g}{(2\pi)^D}.
\end{eqnarray}
At ${\cal O}(\alpha_s)$ the full amplitude is the sum of the amplitudes of the Born term, virtual one-loop
and the real contributions
\begin{eqnarray}
M=M_0+M_{loop}+M_{real}.
\end{eqnarray}
Squaring the full amplitude we have
\begin{eqnarray}
|M|^2=|M_0|^2+|M_{vir}|^2+|M_{real}|^2+{\cal O}(\alpha_s^2),
\end{eqnarray}
where $|M_{vir}|^2=2M_0^\dagger M_{loop}$ and $|M_{real}|^2=M_{real}^\dagger M_{real}$.
Considering Eqs.~\ref{gamma} and \ref{phase}, the virtual corrections to the doubly differential decay rate is then given by
\begin{eqnarray}
\frac{d^2\Gamma_{vir}}{dx_b d\cos\theta_P}=\frac{1-\omega}{32\pi m_t}|M_{vir}|^2\delta(1-x_b),
\end{eqnarray}
where $x_b$ is defined in (\ref{scale}). The one-loop vertex correction and the wave-function renormalization
contain uv- and ir-singularities that  all uv-singularities are canceled after summing all
virtual corrections up whereas the ir-divergences are remaining which are now labeled by $\epsilon$.
Therefore, the virtual doubly differential distribution  reads
\begin{eqnarray}\label{virtualfinal}
\frac{d^2\Gamma_{vir}}{dx_b d\cos\theta_P}=\frac{1}{2}(\frac{d\Gamma^{unpol}_{vir}}{dx_b}+P\frac{d\Gamma^{pol}_{vir}}{dx_b}\cos\theta_P),
\end{eqnarray}
where the unpolarized differential decay rate normalized to the unpolarized Born term is
\begin{eqnarray}
\frac{1}{\Gamma^{unpol}_0}\frac{d\Gamma^{unpol}_{vir}}{dx_b}&=&\nonumber\\
&&\hspace{-1cm}\frac{C_F\alpha_S}{2\pi}\{A-4\frac{1-\omega}{1-4\omega^2}\ln(1-\omega)\}\delta(1-x_b),\nonumber\\
\end{eqnarray}
and the polarized differential width normalized to the polarized Born  width reads
\begin{eqnarray}
\frac{1}{\Gamma^{pol}_0}\frac{d\Gamma^{pol}_{vir}}{dx_b}=\frac{C_F\alpha_S}{2\pi}A\delta(1-x_b),
\end{eqnarray}
with
\begin{eqnarray}
A&=&-\frac{1}{2}\big(2\ln(1-\omega)-\ln\frac{4\pi\mu_F^2}{m_t^2}+\gamma_E-\frac{5}{2}\big)^2+\nonumber\\
&&\frac{1}{\epsilon}\big(2\ln(1-\omega)-\ln\frac{4\pi\mu_F^2}{m_t^2}+\gamma_E-\frac{5}{2}\big)-\nonumber\\
&&\frac{1-4\omega}{1-2\omega}\ln(1-\omega)+2\ln\omega\ln(1-\omega)+2Li_2(1-\omega)\nonumber\\
&&-\frac{1}{\epsilon^2}-5\frac{\pi^2}{12}-\frac{23}{8}.
\end{eqnarray}
Here $\gamma_E=0.5772\cdots$ is the Euler constant and the dilogarithmic function $Li_2(x)$ is defined as
\begin{eqnarray}
Li_2(x)=-\int_0^x\frac{\ln(1-z)}{z}dz.
\end{eqnarray}
As  is seen, the one-loop contribution is purely real. This can be got from an inspection of the one-loop
Feynman diagram Fig.~\ref{feynmandiagrams}b, which does not accept any nonvanishing physical two-particle cut.

\boldmath
\subsection{TREE GRAPH CONTRIBUTION}
\label{real}
\unboldmath

\begin{figure}
\begin{center}
\includegraphics[width=0.7\linewidth]{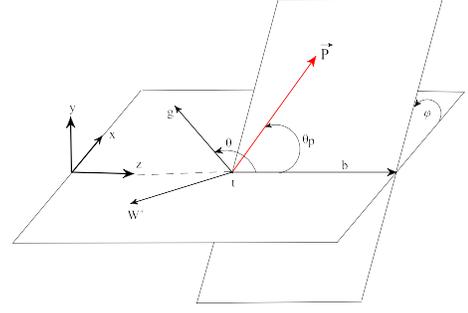}
\caption{\label{feynnlo}%
Definition of  the azimuthal angle $\phi$ and the polar angles $\theta$ and $\theta_P$.
$\vec{P}$ is the polarization vector of the top quark.}
\end{center}
\end{figure}

The ${\cal O}(\alpha_s)$ real graph contribution results from the square of the real gluon emission graphs shown
in Figs.~\ref{feynmandiagrams}(c) and \ref{feynmandiagrams}(d). The real amplitude for the decay process $t(\uparrow)\rightarrow b(p_b)+W^+(p_W)+g(p_g)$ reads
\begin{eqnarray}
M_{real}&=&e g_s\frac{T^n_{ij}}{2\sqrt{2}\sin\theta_W}\bar{u}(p_b, s_b)\{\frac{\gamma^\mu\displaystyle{\not}p_g\gamma^\beta-
2p_t^\beta\gamma^\mu}{2p_t.p_g}+\nonumber\\
&&\frac{\gamma^\beta\displaystyle{\not}p_g\gamma^\mu+
p_b^\beta\gamma^\mu}{2p_b.p_g}\}(1-\gamma_5)u(p_t, s_t)\epsilon^{\star}_\beta(p_g, s_g),\nonumber\\
\end{eqnarray}
where $g_s$ is the strong coupling constant, n is the color index $(n=1,2,3,\cdots ,8)$ so $Tr(T^nT^n)/3=C_F$ and the first and second terms 
in the curly brackets refer to real gluon emission from the top quark
and the bottom quark, respectively. The polarization vector of the gluon is denoted by $\epsilon(p_g, s_g)$.
By working in the massless scheme, the mass of b-quark is set to zero thus the ir-divergences arise from the soft- and collinear
gluon emission. As before to regulate the ir-singularities we work in D-dimensions. In the top quark rest frame (Fig.~\ref{feynnlo})
the momenta and the polarization vector are defined as
\begin{eqnarray}
p_b&=&E_b(1, 0, 0, 1),\nonumber\\
p_g&=&E_g(1, \sin\theta, 0, \cos\theta),\nonumber\\
s_t&=&P(0, \sin\theta_P\cos\phi_P, \sin\theta_P \sin\phi_P, \cos\theta_P).
\end{eqnarray}
Considering (\ref{gamma}) the real differential rate in D-dimensions is given by
\begin{eqnarray}
d\Gamma_{real}&=&\frac{1}{2m_t}\frac{\mu_F^{2(4-D)}}{(2\pi)^{2D-3}}|M_{real}|^2\nonumber\\
&&\hspace{-1cm}\times\frac{d^{D-1}\vec{p}_b}{2E_b}
\frac{d^{D-1}\vec{p}_W}{2E_W}\frac{d^{D-1}\vec{p}_g}{2E_g}\delta^D(p_t-p_b-p_g-p_W).\nonumber\\
\end{eqnarray}
To calculate the  differential decay rate $d\Gamma_{real}/dx_b$ normalized to the Born
width, we fix the momentum of b-quark and integrate over
the energy of gluon which ranges from $E_g^{min}=m_t(1-\omega)(1-x_b)/2$ and  $E_g^{max}=m_t(1-\omega)(1-x_b)/(2(1-x_b(1-\omega)))$
and to obtain the angular distribution of differential width $d^2\Gamma_{real}/(dx_b d\cos\theta_P)$,
 the angular integral in D-dimensions is written as
\begin{eqnarray}
d\Omega_b=\frac{2\pi^{\frac{D}{2}-1}}{\Gamma(\frac{D}{2}-1)}(\sin\theta_P)^{D-4} d\cos\theta_P.
\end{eqnarray}
In the massless scheme, the real and virtual differential widths contain
the poles $\propto 1/\epsilon$ and $1/\epsilon^2$ which disappear only in the total NLO width.
This requires that to get the correct finite terms in the normalized 
doubly differential distribution $1/\Gamma_0\times d^2\Gamma_{real}/(dx_b d\cos\theta_P)$, 
the polarized and unpolarized Born widths will have to be evaluated in dimensional regularization at ${\cal O}(\epsilon^2)$.
We find 
\begin{eqnarray}
\Gamma_0^{unpol}&=&\frac{m_t^3 G_F}{8\sqrt{2}\pi}(1+2\omega)(1-\omega)^2\bigg\{1+\epsilon F+
\epsilon^2 \big[\frac{F^2}{2}+\nonumber\\
&&2\frac{(1+\omega)(1+3\omega)}{(1+2\omega)^2}-\frac{\pi^2}{4}\big]\bigg\},\nonumber\\
\Gamma_0^{pol}&=&\frac{m_t^3 G_F}{8\sqrt{2}\pi}(1-2\omega)(1-\omega)^2\bigg\{1+\epsilon \big[H+\frac{2\omega}{1-2\omega}\big]+\nonumber\\
&&\epsilon^2 \big[\frac{1}{2}(H+\frac{5}{2})^2-H\frac{5-14\omega}{2(1-2\omega)}-\frac{\pi^2}{12}-\frac{25}{8}\big]\bigg\},\nonumber\\
\end{eqnarray}
with 
\begin{eqnarray}
F&=&\ln\frac{4\pi\mu_F^2}{m_t^2}-2\ln(1-\omega)-\gamma_E+2\frac{1+\omega}{1+2\omega},\nonumber\\
H&=&\ln\frac{4\pi\mu_F^2}{m_t^2}-2\ln\frac{1-\omega}{2}-\gamma_E.
\end{eqnarray}
Now the real gluon contribution is given by
\begin{eqnarray}\label{realfinal}
\frac{d^2\Gamma_{real}}{dx_b d\cos\theta_P}=\frac{1}{2}(\frac{d\Gamma^{unpol}_{real}}{dx_b}+P\frac{d\Gamma^{pol}_{real}}{dx_b}\cos\theta_P),\nonumber\\
\end{eqnarray}
where, by defining $T=-\ln(4\pi\mu_F^2/m_t^2)+2\ln(1-\omega)+\gamma_E$, one has
\begin{eqnarray}
\frac{1}{\Gamma^{unpol}_0}\frac{d\Gamma^{unpol}_{real}}{dx_b}&=&
\frac{C_F\alpha_S}{2\pi}\bigg\{\delta(1-x_b)\bigg[\frac{1}{2}T^2-T+\frac{1}{\epsilon^2}-\nonumber\\
&&\frac{2\omega}{1-\omega}\ln\omega+2Li_2(1-\omega)-\frac{1}{\epsilon}(T-1)\nonumber\\
&&-\frac{\pi^2}{4}\bigg]+2(1+x_b^2)\Big(\frac{\ln(1-x_b)}{1-x_b}\Big)_++\nonumber\\
&&\frac{1}{(1-x_b)_+}\bigg[1-4x_b+x_b^2+\nonumber\\
&&\frac{4x_b\omega(1-\omega)(1-x_b)^2}{(1+2\omega)(1-x_b(1-\omega))}+\nonumber\\
&&(1+x_b^2)\bigg(\ln[\frac{x_b^2(1-\omega)^2m_t^2}{4\pi\mu_F^2}]+\nonumber\\
&&\gamma_E-\frac{1}{\epsilon}\bigg)\bigg]\bigg\},\nonumber\\
\end{eqnarray}
and
\begin{eqnarray}
\frac{1}{\Gamma^{pol}_0}\frac{d\Gamma^{pol}_{real}}{dx_b}&=&
\frac{C_F\alpha_S}{2\pi}\bigg\{\delta(1-x_b)\big[\frac{1}{2}T^2-T-\frac{2\omega}{1-\omega}\ln\omega\nonumber\\
&&+2Li_2(1-\omega)-\frac{1}{\epsilon}(T-1)+\frac{1}{\epsilon^2}-\frac{\pi^2}{4}\big]\nonumber\\
&&+2(1+x_b^2)\Big(\frac{\ln(1-x_b)}{1-x_b}\Big)_++\nonumber\\
&&\frac{1}{(1-x_b)_+}\bigg[-1-x_b^2+\frac{8\omega(1-x_b)^2}{1-2\omega}+\nonumber\\
&&4\frac{x_b\omega(1-\omega)(1-x_b)^2}{(1-2\omega)(1-x_b (1-\omega))}+\nonumber\\
&&(1+x_b^2)\bigg(\ln[\frac{x_b^2(1-\omega)^2m_t^2}{4\pi\mu_F^2}]+\gamma_E-\frac{1}{\epsilon}\bigg)\nonumber\\
&&+\frac{8\omega(1-x_b)^2}{x_b(1-\omega)(1-2\omega)}\ln(1-x_b(1-\omega))\bigg]\bigg\}.\nonumber\\
\end{eqnarray}

\boldmath
\section{ANALYTIC RESULTS FOR ANGULAR DISTRIBUTION OF PARTIAL DECAY RATES}
\label{sec:two}
\unboldmath
Now we are in a situation to present our analytic results for the angular distribution
of the differential  decay rate, by summing the Born-level (\ref{bornfinal}),
 the virtual (\ref{virtualfinal}) and real gluon (\ref{realfinal}) contributions.
According to the Lee-Nauenberg theorem, all singularities cancel each other after summing all contributions up and
the final result is free of ir-singularities. Therefore, the complete ${\cal O}(\alpha_s)$
results are
\begin{eqnarray}\label{first}
\frac{d^2\Gamma}{dx_b d\cos\theta_P}=\frac{1}{2}\{\frac{\Gamma^{unpol}_{Nlo}}{dx_b}+P\frac{d\Gamma^{pol}_{Nlo}}{dx_b}\cos\theta_P\},
\end{eqnarray}
that we presented $d\Gamma^{unpol}_{Nlo}/dx_b$  in Ref.~\cite{Kniehl:2012mn} and $d\Gamma^{pol}_{Nlo}/dx_b$ 
in the $\overline{MS}$ scheme is expressed, for the first time, as 
\begin{eqnarray}
\frac{1}{\Gamma^{pol}_0}\frac{d\Gamma^{pol}_{Nlo}}{dx_b}&=&\delta(1-x_b)+
\frac{C_F\alpha_s}{2\pi}\bigg\{\delta(1-x_b)\big[-\frac{3}{2}\ln\frac{\mu_F^2}{m_t^2}\nonumber\\
&&+2\frac{1-\omega}{1-2\omega}\ln(1-\omega)-2\frac{\omega}{1-\omega}\ln\omega-\frac{2\pi^2}{3}\nonumber\\
&&+2\ln\omega\ln(1-\omega)+4Li_2(1-\omega)-6\big]\nonumber\\
&&+2(1+x_b^2)\Big(\frac{\ln(1-x_b)}{1-x_b}\Big)_++\nonumber\\
&&\frac{1}{(1-x_b)_+}\bigg[-1-x_b^2+\frac{8\omega(1-x_b)^2}{1-2\omega}+\nonumber\\
&&4\frac{x_b\omega(1-\omega)(1-x_b)^2}{(1-2\omega)(1-xb (1-\omega))}+\nonumber\\
&&(1+x_b^2)\ln[x_b^2(1-\omega)^2\frac{m_t^2}{\mu_F^2}]+\nonumber\\
&&\frac{8\omega(1-x_b)^2}{x_b(1-\omega)(1-2\omega)}\ln(1-x_b(1-\omega))\bigg]\bigg\}.\nonumber\\
\end{eqnarray}
Since the observed mesons can be also produced through a fragmenting
gluon, therefore, to obtain the most accurate result
for the energy spectrum of meson we have to add
the contribution of gluon fragmentation to the b-quark to produce the outgoing meson. From Fig.~\ref{fig1}, 
it is seen that the contribution of gluon decreases the size of decay rate up to $40\%$ at the threshold,
thus this contribution can be important at low energy of the observed meson. Therefore, 
 the differential decay rate $d\Gamma/dx_g$ is also required  where $x_g$ is defined as $x_g=2 E_g/(m_t(1-\omega))$ as in (\ref{scale}).
To obtain the doubly differential distribution $d^2\Gamma/(dx_g d\cos\theta_P)$, we integrate over the b-quark energy by fixing the
gluon momentum in the phase space. The result is
\begin{eqnarray}\label{second}
\frac{d^2\Gamma}{dx_g d\cos\theta_P}=\frac{1}{2}\{\frac{\Gamma^{unpol}_{Nlo}}{dx_g}+P\frac{d\Gamma^{pol}_{Nlo}}{dx_g}\cos\theta_P\},
\end{eqnarray}
where $d\Gamma^{unpol}_{Nlo}/dx_g$ can be found in our previous work \cite{Kniehl:2012mn} and $d\Gamma^{pol}_{Nlo}/dx_g$ is listed here
\begin{eqnarray} 
\frac{1}{\Gamma^{pol}_0}\frac{d\Gamma^{pol}_{Nlo}}{dx_g}&=&\frac{C_F\alpha_s}{2\pi}\bigg\{
\frac{2}{x_g^2}\ln(1-x_g(1-\omega))\big[\nonumber\\
&&\frac{(1-2x_g)^2}{1-\omega}+\frac{4x_g(1-x_g)}{1-2\omega}\big]+\nonumber\\
&&\frac{1}{2(1-2\omega)}\big[2(1-2\omega)+\frac{4(1+\omega)}{1-\omega}-\nonumber\\
&&x_g(1+6\omega)-\frac{\omega(6\omega^2+\omega+2)}{(1-\omega)(1-x_g(1-\omega))}\nonumber\\
&&-\frac{\omega^2(1-2\omega)}{(1-\omega)(1-x_g(1-\omega))^2}\big]+\nonumber\\
&&\frac{1+(1-x_g)^2}{x_g}\big[2\ln(x_g(1-x_g))-\ln\frac{\mu_F^2}{m_t^2}\nonumber\\
&&-\ln(1-x_g(1-\omega))+2\ln(1-\omega)]\bigg\}.\nonumber\\
\end{eqnarray}

\section{NONPERTURBATIVE FRAGMENTATION AND HADRON LEVEL RESULTS}
\label{sec:three}

\begin{figure}
\begin{center}
\includegraphics[width=1\linewidth,bb=37 192 552 629]{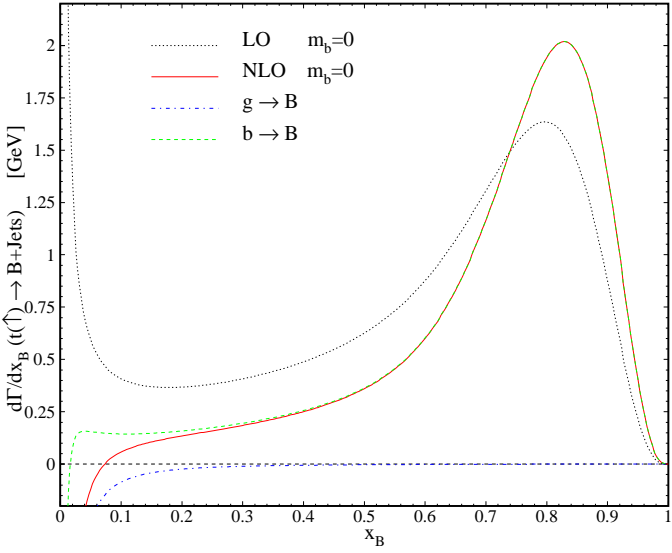}
\caption{\label{fig1}%
$d\Gamma(t(\uparrow)\to BW^++X)/dx_B$ as a function of $x_B$ in the ZM-VFN ($m_b=0$)
scheme. The NLO result (solid line) is compared to the LO one (dotted line) and broken
up into the contributions due to $b\to B$ (dashed line) and $g\to B$
(dot-dashed line) fragmentation. We set $\mu_F=m_t$ and $\mu_{0F}=m_b$.}
\end{center}
\end{figure}

\begin{figure}
\begin{center}
\includegraphics[width=1\linewidth,bb=37 192 552 629]{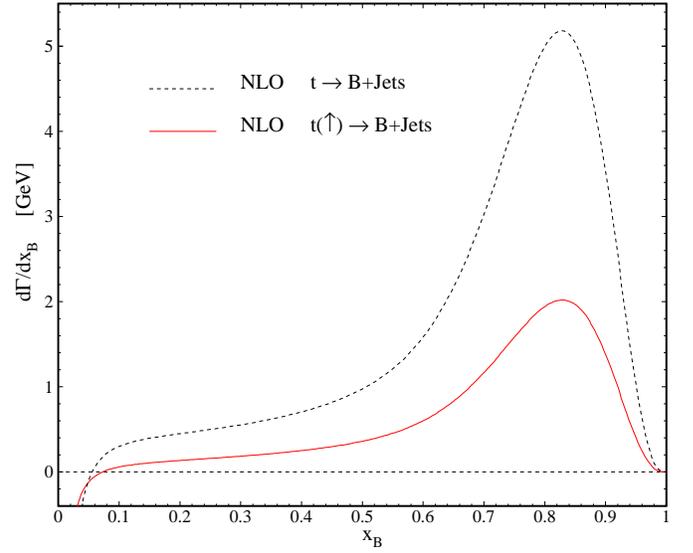}
\caption{\label{fig2}%
$d\Gamma/dx_B$ as a function of $x_B$ in the ZM-VFN ($m_b=0$)
scheme considering the unpolarized (dashed line)
and polarized (solid line) partial decay rates at NLO.}
\end{center}
\end{figure}

\begin{figure}
\begin{center}
\includegraphics[width=1\linewidth,bb=37 192 552 629]{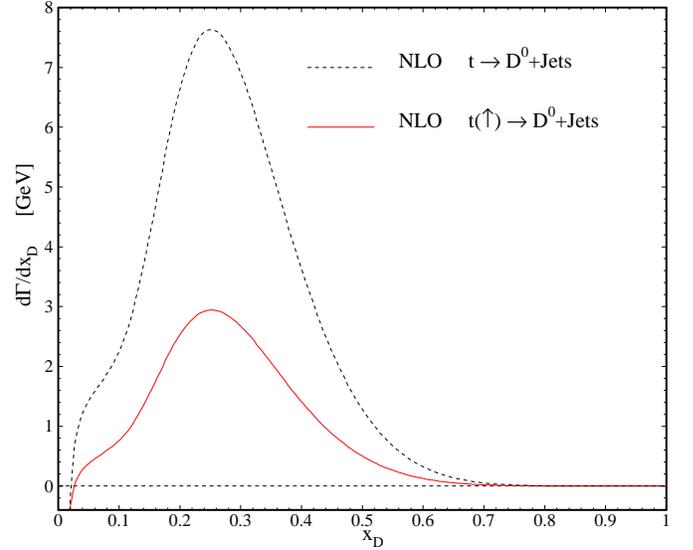}
\caption{\label{fig3}%
$x_D$ spectrum in top decay, with the hadronization
modeled according to the Bowler model considering the unpolarized (dashed line)
and polarized (solid line) decay rates at NLO. We set $\mu_F=m_t$ and $\mu_{0F}=m_b$.}
\end{center}
\end{figure}

\begin{figure}
\begin{center}
\includegraphics[width=1\linewidth,bb=37 192 552 629]{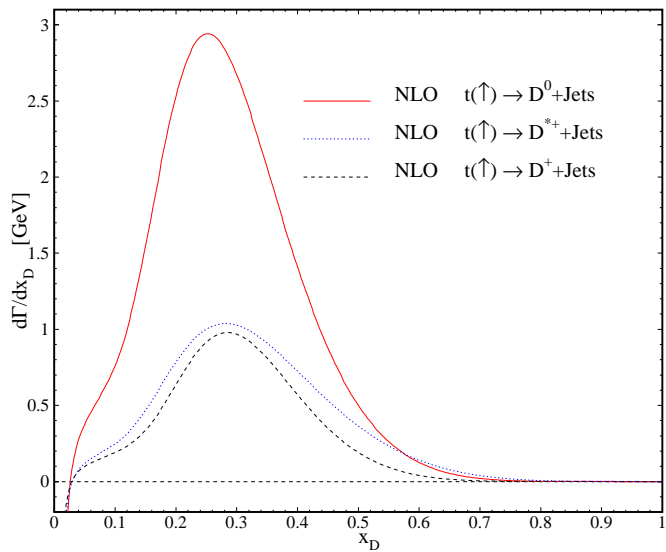}
\caption{\label{fig4}%
$d\Gamma(t(\uparrow)\to DW^++X)/dx_D$ as a function of $x_D$ at NLO 
for $D=D^0$ (solid line), $D=D^{\star+}$ (dotted line) and $D=D^+$ (dashed line).}
\end{center}
\end{figure}

In this section, performing a numerical analysis we present our phenomenological results for the energy spectrum of the
heavy mesons B and D from polarized top decays and compare them  with the 
unpolarized one in \cite{Kniehl:2012mn}. 
We define the normalized-energy fractions of the outgoing mesons similarly
to the parton-level one in (\ref{scale}) as
$x_B=2E_B/(m_t(1-\omega))$ for the B-meson and $x_D=2E_D/(m_t(1-\omega))$
for the D-meson. Considering the factorization theorem of the 
QCD-improved parton model \cite{jc}, the energy distribution of a hadron
can be expressed as the convolution of the parton-level spectrum with the
nonperturbative fragmentation function $D_a^H(z,\mu_F)$ 
\begin{eqnarray}\label{convolute}
\frac{d\Gamma}{dx_H}=\sum_{i=b, g}\frac{d\Gamma}{dx_i}(\mu_R, \mu_F)\otimes D_i^H(\frac{x_H}{x_i},\mu_F),
\end{eqnarray}
where $H$ stands for $B-$ and $D$-mesons and the allowed $x_H$ ranges are $2m_H/(m_t(1-\omega))\leq x_H\leq 1$.
The integral convolution appearing in (\ref{convolute}) is defined as
\begin{eqnarray}
(f\otimes g)(x)=\int_x^1 dx f(z) g(\frac{x}{z}).
\end{eqnarray}
In (\ref{convolute}), $\mu_F$ and $\mu_R$ are the factorization and the renormalization scales, respectively, and
one can use two different values for these  scales; however, a choice often made consists of setting $\mu_R=\mu_F$ 
and we adopt the convention $\mu_R=\mu_F=m_t$ for our results.
In (\ref{convolute}), $d\Gamma/dx_i$ are the parton-level differential rates presented in 
(\ref{first}) and (\ref{second}) and $D_i^H$ are the nonperturbative FFs describing
the hadronizations $b\rightarrow H$ and $g\rightarrow H$ which are process independent.
Several models have been yet proposed to describe the FFs.
In Ref.~\cite{Kneesch:2007ey}, authors calculated the FFs for $D^0, D^+$ and $D^{\star +}$ mesons by fitting the experimental data from
the BELLE, CLEO, OPAL, and ALEPH collaborations in the modified minimal-subtraction ($\overline{MS}$) factorization scheme.
They have parameterized the $z$ distributions of the $b\rightarrow D^0/D^+/D^{\star +}$ FFs at their starting scale $\mu_0=m_b$, as
suggested by Bowler \cite{Bowler}, as
\begin{eqnarray}\label{bw}
D_b^D(z,\mu_0)=Nz^{-(1+\gamma^2)}(1-z)^a e^{-\gamma^2/z},
\end{eqnarray}
while the FF of the gluon is set to zero and this FF is evolved to higher scales using the DGLAP equations \cite{dglap}.
As in \cite{Kneesch:2007ey} is claimed, this parametrization yields the best fit to the BELLE
data \cite{Seuster:2005tr} in a comparative analysis using the Monte-Carlo event generator JETSET/PYTHIA.
The values of fit parameters together with the  achieved values of $\overline{\chi^2}$ are presented in Table~\ref{fit}.
\begin{table}[t]
\caption{\label{fit} Values of fit parameters for $b\rightarrow D^0$, $b\rightarrow D^+$ and $b\rightarrow D^{\star+}$  FFs
at the starting scale $\mu_0=m_b$ resulting 
from the global fit in the ZM
approach together with the   values of $\overline{\chi^2}$ achieved.}
\begin{tabular}{ccccc}
\hline
 &$N$ & $a$ & $\gamma$ &$\overline{\chi^2}$ \\
\hline
$D^0$ &$ 80.8 $&$5.77$& $1.15$&$4.66$
\\
$D^+$ & $163$&$6.93$&$1.40$&$2.21$
\\
$D^{\star+}$ &$14.9$&$3.87$& $1.16$&$7.64$
\end{tabular}
\end{table}
From Ref.~\cite{Kniehl:2008zza} we employ the $b\rightarrow B$ FF  determined at NLO
in the ZM-VFN approach through a global fit to
$e^+e^-$-annihilation data taken by ALEPH \cite{Heister:2001jg} and OPAL
\cite{Abbiendi:2002vt} at CERN LEP1 and by SLD \cite{Abe:1999ki} at SLAC SLC.
The power ansatz $D_b^B(z,\mu_0)=Nz^\alpha(1-z)^\beta$
was used as the initial condition for the $b\to B$ FF at
$\mu_0=m_b$, while the gluon FF was  generated
via the DGLAP evolution.
The fit parameters are $N=4684.1$, $\alpha=16.87$, and $\beta=2.628$ with 
$\overline{\chi^2}=1.495$. Following Ref.~\cite{Nakamura:2010zzi}
we adopt the input values 
$G_F = 1.16637\times10^{-5}$~GeV$^{-2}$,
$m_t = 172.9$~GeV,
$m_b = 4.78$~GeV, 
$m_W = 80.339$~GeV, 
$m_B=5.279$~GeV,
$m_D=1.87$~GeV,
and $\Lambda_{\overline{MS}}^{(5)}=231$~MeV with $n_f=5$ active quark flavors and adjusted such that 
$\alpha_s^{(5)}(m_Z=91.18)=0.1184$.\\
To study the scaled-energy ($x_B$ and $x_D$) distributions of the bottom- and charmed-flavored hadrons
produced in the polarized top decay, we consider the quantities
 $d\Gamma(t(\uparrow)\to B+X)/dx_B$ and $d\Gamma(t(\uparrow)\to D+X)/dx_D$ in the ZM-VFN scheme. 
In Fig.~\ref{fig1}, our prediction for the B-meson  is shown by studying the size of the  NLO 
corrections, by comparing the LO (dotted line) and NLO (solid line) results, and the relative importance of the $b\rightarrow B$ (dashed line)
 and $g\rightarrow B$ (dot-dashed line) fragmentation channels at NLO. We evaluated the LO
 result using the same NLO FFs. The $g\rightarrow B$ contribution is negative and appreciable only
 in the low-$x_B$ region. For higher values of $x_B$, as is expected \cite{Corcella:2001hz},  the NLO result is
 practically exhausted by the $b\rightarrow B$ contribution.
 Note that the contribution of the gluon can not be discriminated. It is calculated to see where
it contributes to $d\Gamma/dx_B$. So this part of the
paper is of more theoretical relevance. In the scaled-energy of mesons as a experimental quantity, all contributions
including the b quark, gluon and light quarks contribute.
In Fig~\ref{fig2}, the scaled-energy ($x_B$) distribution of B-hadrons produced in unpolarized (dashed line)
and polarized (solid line) top quark decays at NLO are studied. As is seen, 
in  the unpolarized top decay the partial decay width at Hadron-level is around
$38\%$ higher than the one in the polarized top decay in the peak region. In Fig.~\ref{fig3}
the same comparison is also down for the transition $b\rightarrow D^0$ applying the Bowler 
model (\ref{bw}) for the FFs. Fig.~\ref{fig3} shows  that the probability to 
produce the charmed-flavored mesons through top quark decays in the high-$x_D$ range ($0.7\alt x_D$) is zero. 
In fig.~\ref{fig4}  we study   the scaled-energy ($x_D$) distribution of charmed-flavored hadrons 
produced in polarized top decays as $t(\uparrow)\rightarrow D^0+X_1$ (solid line), 
$t(\uparrow)\rightarrow D^{\star+}+X_2$ (dotted line) and $t(\uparrow)\rightarrow D^++X_3$ (dashed line). 
Note that our results are valid just for $x_D\geq 2E_D/(m_t(1-\omega))=0.028$ and $x_B\geq 0.078$.

\section{Conclusions}
\label{sec:four}

The top quark decays rapidly so that it has no time to hadronize and passes on its full spin
information to its decay products. The CERN  LHC, as a superlative top
factory, allows us to study top quark decays that within the SM are completely dominated by 
the mode $t\rightarrow W^++b$, followed by $b\rightarrow H+X$. Therefore,
 the distribution in the scaled H-hadron energy $x_H$ in the top
rest frame are of particular interest. In fact, the $x_H$ distribution provides direct access 
to the H-hadron FFs, and its  $\cos\theta_P$ distribution  allows one to  analyze the top quark polarization where
 the polar angle $\theta_P$ refers to the angle between the polarization vector of the top and the z-axis.\\
In \cite{Kniehl:2012mn} we have studied the scaled-energy ($x_B$) distribution of the  B-meson in
unpolarized top quark decays and in the present work we made our predictions for the 
scaled-energy ($x_B, x_D$) distributions of the  B- and D-mesons in the polarized 
top quark rest frame by studying the quantity $d^2\Gamma/(dx_H d\cos\theta_P)$. 
As  was mentioned, the scaled-energy distribution of hadron enables us to deepen our knowledge of the hadronization process and
to pin down the $b, g\rightarrow B/D$ FFs while the angular analysis of the polarized top decay constrain 
these FFs even further. Furthermore, the polarization state of top quarks
can be specified from the angular distribution of the outgoing hadron energy.
The universality and scaling violations of the B- and D-hadron FFs will be able to test at LHC by comparing our NLO predictions with future measurements 
of $d\Gamma/dx_H$ and $d\Gamma(\uparrow)/dx_H$. One can also test the SM and/or non-SM couplings through polarization measurements
involving top quark decays (mostly $t(\uparrow)\rightarrow b+W^+$). 
The formalism made here is also applicable to the production of hadron species
other than B and D hadrons, such as pions and kaons, through the polarized top quark decay using the $b, g\rightarrow \pi/K$ FFs 
presented in our recently paper \cite{maryam}.

\begin{acknowledgments}
I would like to thank Professor Bernd A. Kniehl and Gustav Kramer to propose this topic
 and I would also like to thank Dr Z. Hamedi for reading and improving the english manuscript.
\end{acknowledgments}

\end{document}